# Nonlinear planar Hall effect

Pan He[1†], Steven S.-L. Zhang[2,3†], Dapeng Zhu[1], Shuyuan Shi[1], Olle G. Heinonen[2], Giovanni Vignale[3] and Hyunsoo Yang[1*]

[1]*Department of Electrical and Computer Engineering, and NUSNNI, National University of Singapore, 117576, Singapore*
[2]*Materials Science Division, Argonne National Laboratory, Lemont, Illinois 60439, USA*
[3]*Department of Physics and Astronomy, University of Missouri, Columbia Missouri 65211, USA*

[†]These authors contributed equally to this work. *e-mail: <u>eleyang@nus.edu.sg</u>

**An intriguing property of three-dimensional (3D) topological insulator (TI) is the existence of surface states with spin-momentum locking, which offers a new frontier of exploration in spintronics. Here, we report the observation of a new type of Hall effect in a 3D TI $Bi_2Se_3$ film. The Hall resistance scales linearly with both the applied electric and magnetic fields and exhibits a $\pi/2$ angle offset with respect to its longitudinal counterpart, in contrast to the usual angle offset of $\pi/4$ between the linear planar Hall effect and the anisotropic magnetoresistance. This novel nonlinear planar Hall effect originates from the conversion of a nonlinear transverse spin current to a charge current due to the concerted actions of spin-momentum locking and time reversal symmetry breaking, which also exists in a wide class of non-centrosymmetric materials with a large span of magnitude. It provides a new way to characterize and utilize the nonlinear spin-to-charge conversion in a variety of topological quantum materials.**

Three-dimensional (3D) topological insulators (TIs) have attracted broad interests in various areas of condensed matter physics, such as spintronics, quantum computing, superconductivity etc. [1-5]. A unique feature of this class of novel materials is the existence of Dirac surface states protected by band topology. The spins of the carriers in these topological surface states (TSS) are locked perpendicularly to their momenta – an intriguing phenomenon known as spin-momentum locking [6-9]. Highly efficient magnetic switching [10-14] and spin-to-charge conversion [15-20] have been demonstrated by taking advantage of the spin-momentum locked TI surface states, which hold great promise for applications in future spintronic devices.

The TI surface states have been shown to be intimately related with a variety of peculiar magnetotransport properties including novel linear and nonlinear magnetoresistance effects in nonmagnetic TI films [21] and bilayer structures consisting of a ferromagnetic layer and a nonmagnetic TI layer [22,23]. Although the magnetoresistance effects in TIs have been investigated intensively, only a few studies have recently reported on the Hall effect associated with TSS. A linear planar Hall effect was discovered in nonmagnetic TI $Bi_{2-x}Sb_xTe_3$ thin films [24], which was attributed to the anisotropic backscattering induced by an in-plane magnetic field and a nonlinear Hall effect due to asymmetric magnon scattering was observed in a magnetic-TI/nonmagnetic-TI bilayer [25].

In this Letter, we report a new type of nonlinear planar Hall effect in a prototypical 3D nonmagnetic TI $Bi_2Se_3$ film, for which the nonlinear Hall resistance is linearly proportional to both the applied electric and magnetic fields, in contrast to other types of Hall effects observed previously in TI systems. The nonlinear planar Hall effect arises from the generation of a transverse pure spin current occurring at the second order of the electric field due to symmetric distortion of the 2D Fermi contour, which can be converted to a nonlinear Hall current by applying an in-plane magnetic field collinear with the spin orientation of the transverse nonlinear spin current. In addition, we discuss the universality of the nonlinear planar Hall effect in other non-centrosymmetric materials including a Weyl semimetal $WTe_2$ and a perovskite oxide $SrTiO_3$ based two dimensional electron gas (2DEG).

In our experiments, high-quality $Bi_2Se_3$ films were grown on $Al_2O_3$ (0001) substrates in a molecular beam epitaxy system with a base pressure $< 2\times10^{-9}$ mbar using the two-step deposition procedure [26]. For the transport measurements, a capping layer of MgO (2 nm)/$Al_2O_3$ (3 nm) was deposited on top of the $Bi_2Se_3$ films as a protection layer, as shown schematically in Fig. 1a. Hall bar devices with channel length $L = 100$ μm and the width $W = 20$ μm (Fig. 1b,c) were fabricated using the standard photolithography and Ar ion milling. Details of the device fabrications of $SrTiO_3$(001) and $WTe_2$ can be found in Supplemental Material [27]. To explore the nonlinear Hall effect, we performed ac harmonic Hall voltage measurements in a physical property measurement system (Quantum Design) with lock-in techniques [43]. Here, we focus on the second harmonic Hall voltage generation under an ac current and in-plane magnetic field, which is associated with the nonlinear transverse transport in response to the second-order electric field [44]. The experimental configuration is different from that of conventional thermopower measurements, wherein an ac current is used to induce an in-plane temperature gradient and the thermoelectric voltage is measured under zero magnetic field [45].

Up to the second order in the applied current $I_x$ along the *x*-direction, the Hall voltage can be expressed as $V_y = R_{yx}^{(1)} I_x + R_{yx}^{(2)} I_x^2$ where both coefficients $R_{yx}^{(1)}$ and $R_{yx}^{(2)}$ are independent of the current. The first term on the right-hand-side is linearly proportional to the current $I_x$ (or electric field $E_x$), which may include the contributions from the ordinary Hall effect (OHE) and a linear planar Hall effect [24]; the second term, proportional to the second order of $I_x$, describes the second harmonic Hall voltage generation $V_y^{2\omega} = \frac{1}{2} R_{yx}^{(2)} I^2 \sin(2\omega t - \pi/2)$, when an ac current $I^\omega = I \sin\omega t$ is applied with the amplitude $I$ (see Supplemental Material [27]). The nonlinear planar Hall effect in a Bi$_2$Se$_3$ film was detected via the second harmonic Hall voltage with an external magnetic field in the film plane forming an angle φ with the longitudinal current, as shown in Fig. 1c. For a fixed current and magnetic field, the measured second harmonic resistance $R_{yx}^{2\omega} (\equiv V_y^{2\omega} / I)$ exhibits a cosine angular dependence (Fig. 1d). It vanishes when the magnetic field and the current are orthogonal (i.e., φ = 90° or 270°), and reaches a minimum (maximum) when magnetic field is parallel (antiparallel) to the current direction. Note the nonlinear Hall effect previously observed in a magnetic/nonmagnetic-TI heterostructure [25] required a magnetic layer to provide asymmetric magnon-mediated scattering. This mechanism is not applicable to our nonmagnetic Bi$_2$Se$_3$ film due to the absence of electron-magnon scattering. The nonlinear planar Hall effect in this work arises intrinsically from the topological surface band structure, as we discuss below.

We further measured $R_{yx}^{2\omega}$ for different magnitudes of *I* and *H*. As shown in Fig. 1e, for a fixed *H* = 2 *T*, the amplitude of the φ-dependent $R_{yx}^{2\omega}$ ($\Delta R_{yx}^{2\omega}$), exhibits a linear dependence on the magnitude of the current *I*, manifesting the nonlinear nature of $R_{yx}^{2\omega}$ [46,47]. $\Delta R_{yx}^{2\omega}$ is also found to scale linearly with the magnitude of *H*, as shown in Fig. 1f. The linear *H*-dependence of $R_{yx}^{2\omega}$ is a distinctive feature of our nonlinear planar Hall effect in the single layer TI thin film and qualitatively different from that in magnetic/nonmagnetic-TI bilayer – the latter decreases with increasing *H* for a saturated magnetic layer due to the suppression of magnon scattering [25]. The observed nonlinear Hall effect (Fig.1d-f) in $Bi_2Se_3$ films takes the form of $R_{yx}^{2\omega}$ ~ **E·H**, where **E** is the applied electric field and **H** is the magnetic field applied in the plane of the films.

In addition to the second harmonic transverse resistance $R_{yx}^{2\omega}$, we have measured its longitudinal counterpart $R_{xx}^{2\omega}$ (Fig. 2a) for the same sample – the recently observed BMER effect [21,48], which linearly scales with both the electric and magnetic fields as well. A comparison of the angular dependences of the two nonlinear resistances shown in Fig. 2b reveals a 90° angle offset, in contrast to the usual 45° angle difference which exists between the linear transverse planar Hall effect and the longitudinal anisotropic magnetoresistance in TI thin films [24] as well as polycrystalline ferromagnetic metal thin films [49] with the external magnetic field or magnetization rotated in the film plane. The 90° angle difference is rooted in different responses of electrons occupying the TSS to the applied electric and magnetic fields. While the distortion of electron distribution (or the Fermi contour) in response to the second order of the electric field is symmetric in **k**-space [21,50], it becomes asymmetric when a magnetic field is applied. More specifically, when the magnetic field **H** is applied in the *x*-direction, the Fermi contour is distorted asymmetrically along the *y*-direction (orthogonal to **H**), due to the combined actions of spin-momentum locking and the hexagonal warping effect [21] as illustrated schematically in Fig. 2c,d. Such an asymmetric distortion of the Fermi contour, as we will examine in the following theoretical calculations, is reflected in the nonlinear transport, i.e, the nonlinear planar Hall and the BMER effects occurring in the transverse and longitudinal directions, respectively.

The nonlinear planar Hall effect originates from the conversion of a transverse nonlinear spin current to a nonlinear charge Hall current by applying an in-plane magnetic field. To see this, we analyze the spin and charge transport in the TSS described by the model Hamiltonian [51,52], $H_{TI} = \frac{\hbar^2 k^2}{2m^*} + \alpha_D \hbar(\sigma_x k_y - \sigma_y k_x) + \frac{1}{2}\lambda(k_+^3 + k_-^3)\sigma_z + g\mu_B \mathbf{H} \cdot \boldsymbol{\sigma}$, where the first term generates particle-hole asymmetry with effective mass $m^*$, $\alpha_D$ is the Dirac velocity, $\boldsymbol{\sigma}$ denote the Pauli spin matrices, the term cubic in $\mathbf{k}$ describes the hexagonal warping effect with a parameter $\lambda$ [51], and $k_\pm = k_x \pm i k_y$. In the presence of $\mathbf{E}$, the nonequilibrium electron distributions in the second order of $\mathbf{E}$ may be expressed as $\delta f(\mathbf{k}) \sim \frac{\partial^2 f_0}{\partial k_i^2} E_i^2$, where $f_0(\varepsilon_\mathbf{k})$ is the equilibrium Fermi-Dirac distribution with $\varepsilon_\mathbf{k} = \frac{\hbar^2 k^2}{2m^*} + \sqrt{(\alpha_D \hbar k)^2 + [\lambda(k_x^3 - 3k_x k_y^2)]^2}$ for the upper band, which is in general even in $\mathbf{k}$, i.e., $\delta f(-\mathbf{k}) = \delta f(\mathbf{k})$, as schematically shown in Figs. 3a,b; in other words, the nonequilibrium surface states with opposite wavevectors and opposite spins (due to the spin-momentum locking) are equally populated, which leads to a nonlinear spin current $Q_a^b \sim \mathbf{E}^2$, where the superscript $b$ and the subscript $a$ denote the direction of spin flow and the direction of the spin, respectively. Using the Boltzmann transport formulation in the relaxation time approximation (see Supplemental Material [27]), we find the nonzero transverse nonlinear spin current components, $Q_y^x = \frac{\varepsilon_F (e\tau E_x)^2}{64\pi \alpha_D \hbar}\left(\frac{3\lambda^2 \varepsilon_F^3}{\alpha_D^4 \hbar^6} + \frac{1}{m^*}\right)$, when the electric field is applied in the $x$-direction (Fig. 3b), which are related to the hexagonal warping and the particle-hole asymmetry, respectively. By applying an in-plane magnetic field collinear with the spin direction of the spin current, one creates an imbalance between the two spin fluxes of the spin current and partially converts it into the nonlinear planar Hall current (Fig. 3d). Here, we emphasize that such spin-to-charge interconversion would not take place in the absence of hexagonal warping and particle-hole asymmetry. Physically this point may be understood as follows: For TSS with purely linear dispersion, an in-plane magnetic field is equivalent to a shift of the origin of momentum space, which would *not* alter the current, provided the system has translational symmetry in the *x-y* plane.

With the same formulation, the second harmonic Hall resistivity can be expressed as $\rho_{yx}^{2\omega} = -\left(\frac{1}{3}\chi' + \chi''\right)EH\cos\varphi$, where $\chi' = \frac{36\pi g\mu_B \lambda^2 \varepsilon_F}{e\alpha_D^5 \hbar^4}$ and $\chi'' = \frac{g\mu_B \hbar^2}{e\pi\alpha_D m^* \varepsilon_F^2}$. From the ordinary Hall measurement (Supplemental Material [27]), we obtained the carrier concentration $n \sim 7\times 10^{13}$ cm$^{-2}$ and the Fermi energy $\varepsilon_F \sim 400$ meV for which a hexagonal warping effect was observed to be profound [53-55]. Using parameters of $\alpha_D = 5 \times 10^5$ m·s$^{-1}$ [56,57], $\lambda = 50$–128 eV·Å$^3$ [53,54,58], g = 2, $H_x$ = 2 T, $E_x$ = 300 V·cm$^{-1}$, and $m^* = 0.07 m_e$ [59] ($m_e$ is the free electron mass), the second harmonic resistance is estimated to be $R_{yx}^{2\omega} = 0.9 - 2.4$ $m\Omega$. It is in agreement with the measured value of $R_{yx}^{2\omega} = 1.1$ $m\Omega$ at $T = 5$ K. We note that for the above materials parameters, the contributions related to hexagonal warping and particle-hole asymmetry are of the same order of magnitude.

In addition to the TSS, a two-dimensional electron gas (2DEG) with Rashba splitting is known to be present at the surface due to the surface band bending, which may also contribute to the nonlinear planar Hall effect. However, the conventional Rashba 2DEG does not contribute to the nonlinear planar Hall effect, due to an exact cancellation of two subbands with opposite spin chirality (see Supplemental Material [27]). This is a remarkable feature of the nonlinear transport for which only the TSS comes into play – for most linear response effects in TIs both TSS and Rashba 2DEG contribute although usually the TSS is found to play a dominant role [55,60,61]. It turns out that the sign of the observed effect in Fig. 2b follows directly from the spin-momentum locking and the associated nonlinear spin-to-charge-current conversion in the top TSS, as illustrated in Figs. 2 and 3. To obtain a full quantitative understanding, the nondegenerate top and bottom topological surfaces need to be taken into account. The film-thickness dependence with a sign reversal of the nonlinear planar Hall effect indicates that it is indeed a surface effect rather than a bulk effect (see Supplemental Material [27]).

Furthermore, the angle-relation between the nonlinear transverse and longitudinal resistances shown in Fig. 2b is also expected theoretically. The longitudinal second harmonic resistivity with an in-plane magnetic field was derived as $\rho_{xx}^{2\omega} = \chi' \hat{\mathbf{z}} \cdot (\mathbf{E} \times \mathbf{H}) = \chi' EH \sin\varphi$ by assuming the hexagonal warping in the nonlinear transport [21]. Comparing with $\rho_{yx}^{2\omega}$ one can identify an angle difference of $\pi/2$. In addition, we note that, given the magnitudes of the electric and magnetic field, the peak-values of $\rho_{yx}^{2\omega}$ and $\rho_{xx}^{2\omega}$ in the in-plane field scan differ by a factor of 1/3 which arises from hexagonal warping due to the $C_{3v}$ symmetry of the TI about the trigonal $z$-axis [51]. The factor of 1/3 may play an important role in distinguishing the nonlinear planar Hall from the Nernst effect due a possible thermal gradient $\nabla T$ perpendicular to the TI layer. The Nernst effect, if present, would produce a voltage proportional to $(\mathbf{H} \times \nabla T)$, which has the same angular dependence as observed for the second harmonic resistances. The contribution of the possible Nernst effect to the nonlinear resistivity, however, does not depend on the direction of the electric field, and hence should be isotropic in the in-plane magnetic field scan. In other words, the Nernst effect should give rise to the same nonlinear resistivity in the transverse and longitudinal directions [46,62], i.e., $\rho_{yx}^{2\omega} = \rho_{xx}^{2\omega}$. For a direct comparison, we convert the measured resistance to resistivity and obtain $\frac{\rho_{yx}^{2\omega}}{\rho_{xx}^{2\omega}} = \frac{R_{yx}^{2\omega}}{R_{xx}^{2\omega}} \cdot \frac{L}{W}$, where $L$ and $W$ are the length and width of the current channel, respectively in Fig. 1c. We find the measured $\rho_{yx}^{2\omega}/\rho_{xx}^{2\omega}$ is about 0.22, which is much less than unity but close to our theoretical value 1/3. This indicates that the Nernst effect is *not* the dominant mechanism for the measured nonlinear resistances in our TI samples.

In addition to the nonlinear planar Hall effect, we have detected a nonlinear Hall signal in the out-of-plane field scan. However, the measured second harmonic Hall resistance does not exhibit a regular sinusoidal angular dependence nor a linear dependence on the magnetic field [27], in contrast to the theoretical out-of-plane nonlinear Hall originating from the conversion of the nonlinear spin current with out-of-plane spin component [27]. One possible reason for this unexpected behavior is the contribution from the classical Lorentz force (i.e., $-\frac{e}{c}\mathbf{v} \times \mathbf{H}$) which is inoperative for planar Hall effect [27].

As a final point, we expect a similar nonlinear planar Hall effect to exist in a broad range of noncentrosymmetric materials with strong spin orbit coupling, whereby the spin polarized bands would give rise to a nonlinear transverse spin current in nonequilibrium conditions [63] which may be converted to nonlinear charge Hall currents by breaking the time reversal symmetry. For example, we have found the nonlinear planar Hall effect in the 2DEG on the surface of a perovskite oxide $SrTiO_3$(001) [64] and in the transition metal dichalcogenide $WTe_2$ [65] with a peculiar spin texture [66] (see Supplemental Material [27]). In Table I, the nonlinear Hall resistivity (conductivity) per unit electric and magnetic fields $\rho_{yx}^{2\omega}/(E_x H_y)$ [$\sigma_{yx}^{2\omega}/(E_x H_y)$] for different materials, is summarized for comparison. The low crystal symmetry in WTe$_2$ and the high carrier mobility in SrTiO$_3$ 2DEG are favorable factors for generating a larger nonlinear planar Hall effect. Note the 2DEG on SrTiO$_3$ (001) surface differs from the conventional Rashba 2DEG, as the cubic Rashba effect and multiorbital effects of *d* electron play important roles. Its nonlinear planar Hall effect shows a sign change with temperature, while BMER shows no sign change (see Supplemental Material [27]). This indicates that different contributions to the nonlinear planar Hall effect need to be identified, distinct from those of the BMER. Our observations of the nonlinear planar Hall effect may stimulate future studies of nonlinear transports in a wide variety of materials with nontrivial spin textures. Its potential applications in rectification and second harmonic generations can be further extended to the gigahertz and terahertz ranges.

In conclusion, we have observed a nonlinear planar Hall effect in nonmagnetic TI Bi$_2$Se$_3$ films with two distinctive features: a linear dependence on both the applied electric and magnetic fields, and a $\pi/2$ angle offset relative to its longitudinal counterpart (BMER). Physically, this nonlinear planar Hall effect arises from the conversion of a nonlinear transverse spin to charge current in the presence of a magnetic field collinear with spins, due to the spin-momentum locking. The nonlinear planar Hall effect also occurs in a broad range of noncentrosymmetric materials with strong spin-orbit coupling and nontrivial spin textures. Therefore, we envision our findings pave a new route to the emerging field of nonlinear spintronics.

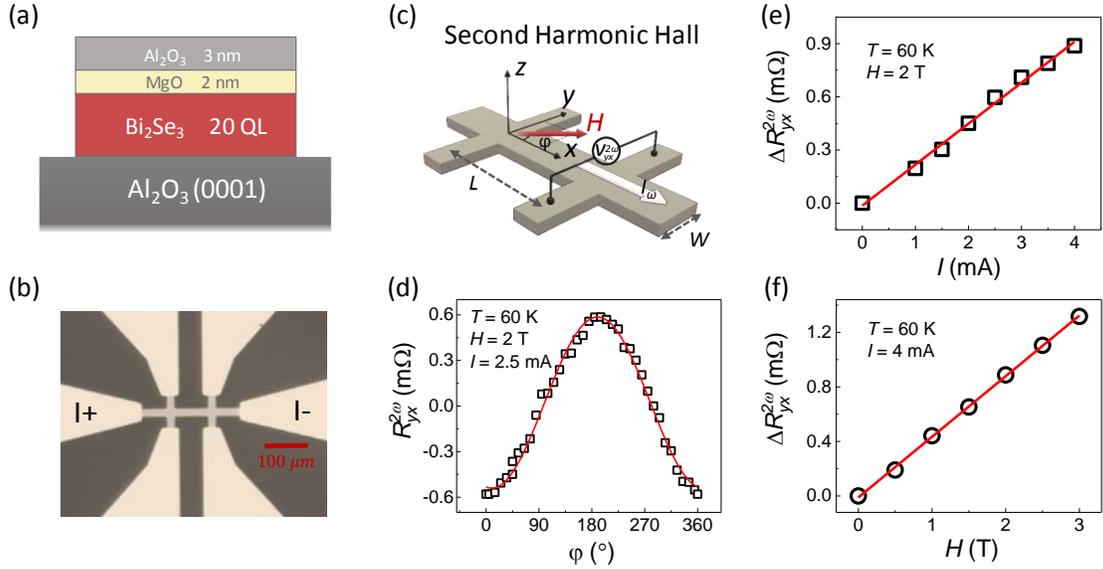

FIG. 1. (a) Schematic illustration of the sample structure. (b) Optical image of a Hall bar. (c) Schematic of the nonlinear planar Hall effect measurements. A sinusoidal current $I_\omega$ with a frequency of 21 Hz was applied to the Hall bar device, and the out-of-phase (-90°) second harmonic transverse voltage $V_{yx}^{2\omega}$ was measured under the presence of an in-plane (*xy*) magnetic field **H**. (d) Angle (φ)-dependence of the nonlinear Hall resistance $R_{yx}^{2\omega}$ while rotating **H** with respect to the current direction in a 20 quintuple layer (QL, 1 QL ≈ 1 nm) Bi$_2$Se$_3$ film. A vertical offset was subtracted for clarity. The solid line is a sinφ fit to the data. A linear dependence of the sinusoidal amplitude $\Delta R_{yx}^{2\omega}$ on the current *I* (e) and magnetic field *H* (f). The solid lines are linear fits.

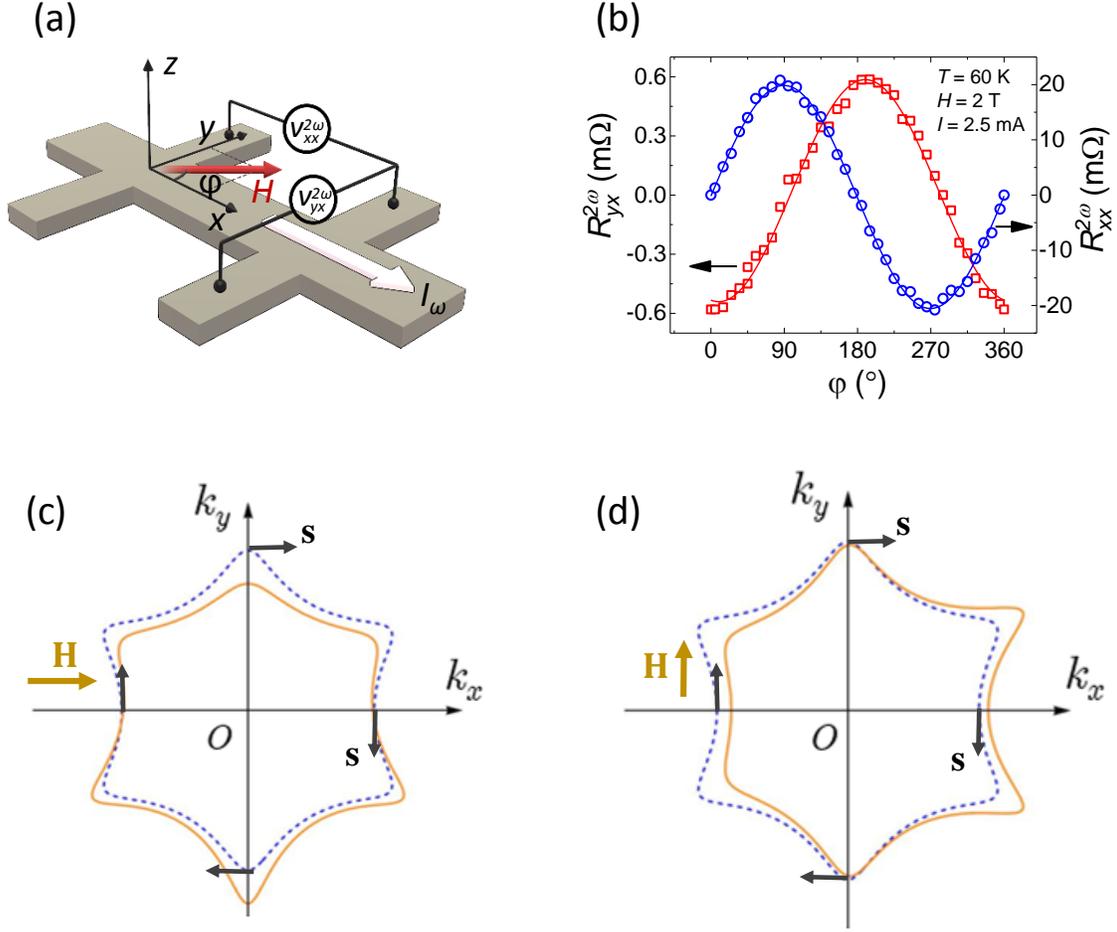

FIG. 2. (a) Schematic illustration of the simultaneous measurements of nonlinear planar Hall effect ($V_{yx}^{2\omega}$) and nonlinear magnetoresistance ($V_{xx}^{2\omega}$). (b) Comparison of the angular dependence of $R_{yx}^{2\omega}$ and $R_{xx}^{2\omega}$ while rotating **H** in-plane in a 20 QL Bi$_2$Se$_3$ film. (c) When **H** is aligned in the *x*-direction, an asymmetric distortion of the Fermi contour is induced along the *y*-direction. (d) When **H** is aligned in the *y*-direction, an asymmetric distortion of the Fermi contour occurs along the *x*-direction. The blue (yellow) curves show the schematic Fermi contours of the surface band under a zero (nonzero) external magnetic field. The black arrows indicate the spin directions of the four typical TSS.

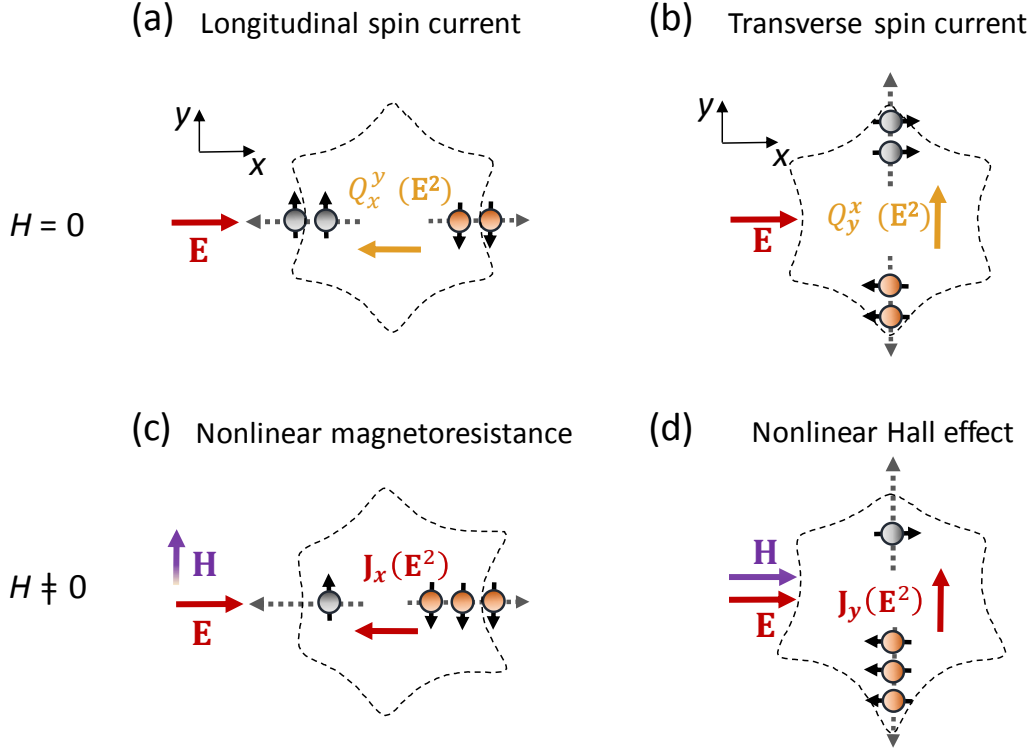

FIG. 3. When an electric field **E** is applied to the 3D TI, a nonlinear spin current $Q_a^b$ at the second order of **E** is generated simultaneously at both the longitudinal $Q_x^y$ (a) and transverse $Q_y^x$ (b) directions, due to the perpendicularly spin-momentum locked surface states. (c) When an external magnetic field **H** is applied perpendicular to **E**, the longitudinal nonlinear spin current is partially converted into a charge current $\mathbf{J}_x(\mathbf{E}^2)$, giving rise to the nonlinear magnetoresistance. (d) When an external magnetic field **H** is applied parallel to **E**, the transverse nonlinear spin current is partially converted into a charge current $\mathbf{J}_y(\mathbf{E}^2)$, giving rise to the nonlinear Hall effect. The gray arrows indicate the **k** directions and the black arrows indicate the spin directions of TI surface state.

TABLE I. Comparison of the nonlinear planar Hall effect for various materials with nontrivial spin textures. Note that the data for WTe$_2$ are obtained when the current is applied along the $a$ and $b$ axis. The thickness of 2DEG on SrTiO$_3$(001) surface were taken from references[67,68].

| Materials | T (K) | $\rho_{xx}$ (μΩ·cm) | W/L (μm/μm) | t (nm) | $R_{yx}^{2\omega}/R_{xx}^{2\omega}$ | $\rho_{yx}^{2\omega}/\rho_{xx}^{2\omega}$ | $\rho_{yx}^{2\omega}/(E_xH_y)$ (mΩ· V$^{-1}$·μm$^2$·T$^{-1}$) | $\sigma_{yx}^{2\omega}/(E_xH_y)$ (mΩ$^{-1}$· V$^{-1}$·T$^{-1}$) |
|---|---|---|---|---|---|---|---|---|
| Bi$_2$Se$_3$ | 5 | 433 | 20/100 | 19 | 0.044 | 0.22 | 0.02 | 0.001 |
| SrTiO$_3$ | 2 | 280 | 20/115 | ~20 | 0.018 | 0.1 | 13.8 | 1.76 |
| WTe$_2$ ($b$) | 2 | 147 | 3/3 | 13.4 | 0.5 | 0.5 | 4.02 | 1.85 |
| WTe$_2$ ($a$) | 2 | 43 | 3/3 | 13.4 | 1.9 | 1.9 | 2.84 | 15.4 |